# Generation of spatiotemporal optical vortices with partial temporal coherence

AMAL MIRANDO, [1, †] YIMIN ZANG, [1, †] QIWEN ZHAN[1‡], ANDY CHONG [1, 2, *]

[1]Department of Electro-Optics and Photonics, University of Dayton, Dayton, Ohio 45469, USA
[2]Department of Physics, University of Dayton, Dayton, Ohio 45469, USA
*Corresponding author: achong1@udayton.edu

**Recently, a spatiotemporal optical vortex (STOV) with a transverse orbital angular momentum (OAM) has been generated from coherent ultrafast pulses using mode-locked lasers. In contrast, we demonstrate theoretically and experimentally that a STOV can be generated from a light source with partial temporal coherence with fluctuating temporal phase. By eliminating the need of mode-locked laser sources, the partially coherent STOV will serve as a convenient and cost-effective transverse OAM source.**

Optical vortex beams are well-established phenomena in modern-day singular optics. As shown in 1992 by Allen et al. [1], optical vortex beams, which can be generated by a variety of techniques, have an orbital angular momentum (OAM) [2-5]. A typical vortex beam has a spatial spiral phase of the form $e^{il\emptyset}$ where $\emptyset$ is the azimuthal angle and $l$ is the topological charge. The vortex beams attracted significant interest leading to many applications such as optical tweezers [6], Quantum information processing [7, 8] and free space optical communication [9].

Partially coherent vortex (PCV) beams have also emerged drawing significant interest due to their unique properties in beam shaping, beam rotation and self-reconstruction [10]. A key feature of the PCV beams is that the phase singularity is diminished due to random phase fluctuations. PCV beams have important applications such as optical trapping where focused PCV beams are utilized for beam shaping. In addition, the self-reconstruction ability of PCVs on propagation has been used for information encryption and decryption [11].

Recently, an optical vortex with a transverse OAM has been generated successfully [12, 13]. Here the spiral phase and phase singularity exist in the spatiotemporal (ST) domain as opposed to optical vortex beams. This new form of vortex is called the ST optical vortex (STOV) and it carries the potential for novel applications such as optical manipulation, second harmonic generation, ST spin-orbit coupling, etc. [12].

Similar to PCVs in the spatial domain, STOVs with partial coherence has also been studied theoretically by M. Hyde [14]. His work was focused on the theoretical descriptions of the propagation of partially coherent vortex in the space-frequency and ST domains. Hyde suggests that these partially coherent vortices will be useful in the fields of optical tweezing, particle manipulation, optical communications, and astronomy where optical vortices are currently utilized. Thus, further experimental study of STOVs with partial coherence will be an interesting direction.

Here, we present a simple method to generate a STOV using a light source with partial temporal coherence. The experimental results are well verified by numerical simulations. Since cheaper broadband sources such as light-emitting diodes can be used in the place of expensive mode-locked lasers, this work provides cheap STOV sources for future applications.

Due to the conservation of the OAM under the Fourier transform, a simple and reliable method can generate the STOV [12]. The same concept is applied to partial temporal coherent STOV generation with partially coherent sources such as amplified spontaneous emission (ASE), where the degree of temporal coherence is reduced by the randomly emitted photons [15, 16]

ASE can be numerically simulated by applying a randomly distributed spectral phase [17, 18]. To model the ASE, a Gaussian spectrum with a randomly distributed spectral phase is simulated. The corresponding pulse $A(t)$ can be calculated by the Fourier transform, which is denoted by equation 1.

$$A(t) = FT\{e^{-\omega^2/r^2} e^{j\phi_{rand}(\omega)}\}, \quad \phi_{rand}(\omega) \sim N(\mu, \sigma^2) \quad (1)$$

In equation 1, $\phi_{rand}(\omega)$ represents the random spectral phase which is a normal distribution, $N(\mu, \sigma^2)$, where μ is the mean value and $\sigma^2$ is the variance. Obviously, $\sigma^2 = 0$ represents the mode-locked pulse. Larger noise phase fluctuation is modeled by larger $\sigma^2$. Figure 1 shows some temporal profiles with different amounts of $\sigma^2$. By increasing the $\sigma^2$, the temporal profile changes from a clean mode-locked pulse to a partial temporal coherent random pulse.

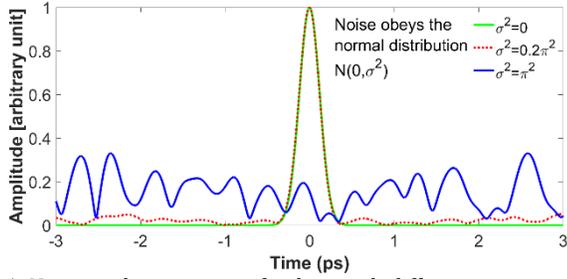

Fig. 1. Numerical generation of pulses with different noise spectral phase (the amplitude of each pulse is adjusted to show fine structures).

Now we can apply the spiral phase $e^{il\phi}$ on the spatial-frequency and frequency domain of the partial temporal coherent pulse. By the two-dimensional Fourier transform, the STOV with partial temporal coherence is numerically calculated. To be consistent with the experimental condition, a Gaussian spectrum is used with the center wavelength of 1030 nm and full width of half maximum bandwidth ~8 nm. Figure 2 shows the amplitude of the simulated STOVs with a topological charge $l$=1. As the phase randomness increases, the shape of the STOV is severely distorted from the ring-shaped profile with multiple singularities occurring at various temporal locations. The presented simulation results are only a small portion of the entire random pulse to reveal the characteristics of the partially coherent STOV. Amplitude fluctuations randomly spread throughout the whole temporal domain with random multiple amplitude peaks.

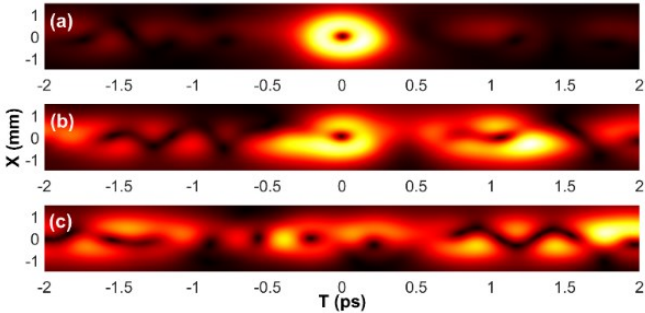

Fig.2. Amplitude of partial temporal coherent STOV with topological charge $l$ = 1, variance of each normal distribution is: (a) $\sigma^2 = 0.2\pi^2$, (b) $\sigma^2 = 0.5\pi^2$, (c) $\sigma^2 = \pi^2$.

To visualize the phase and the corresponding Poynting vectors, the time averaged Poynting vector <**S**> is calculated by equation (2)

$$< S > \propto i(u\nabla u^* - u^*\nabla u)/2 + k|u|^2 \mathbf{z} \qquad (2)$$

where $u$ is the amplitude of the wave packet and $k$ is the wave vector [1]. Figure 3 shows the phase of the partial temporal coherent STOV and corresponding time-averaged Poynting vectors. In equation 2, since the second term is proportional to the linear momentum, we only plot the first term to reveal complex spiral phase structure on the ST domain. Obviously, the simulated Poynting vectors and the corresponding phase map indicate the existence of multiple singularities of the partially coherent STOV.

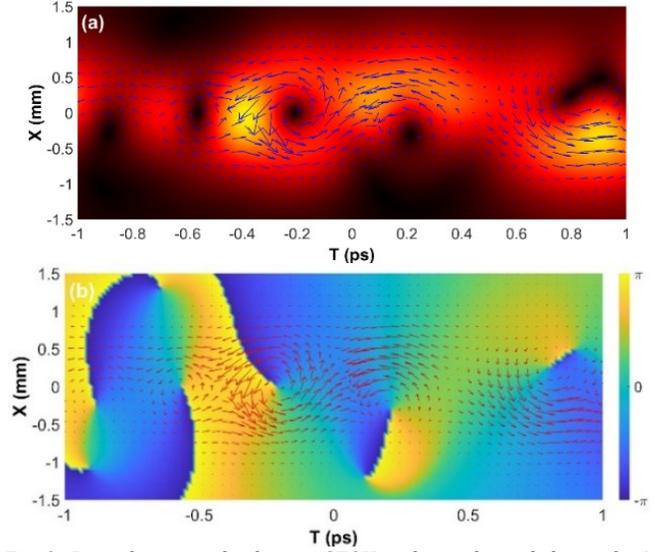

Fig. 3. Partial temporal coherent STOV with topological charge $l$ = 1, $\sigma^2 = \pi^2$, (a) electric field amplitude with Poynting vectors; (b) phase and Poynting vectors.

Previous research has demonstrated that a STOV with a higher topological charge $l > 1$ is not stable under propagation due to the ST astigmatism [12]. For the partially coherent STOV, the random phase itself acts as the ST astigmatism, thus, the higher-order STOV will break up into multiple singularities even without propagation. The instability of higher order partial temporal coherent STOVs is analogous to the spatial vortex splitting due to the atmospheric turbulence [19].

Figure 4(a) shows the partial temporal coherent STOV with $l$ = 2 without propagation. It is clear that the singularity with the topological charge $l$ = 2 splits into two fundamental vortices due to the random phase, which is shown in Figures 4(b) and 4(c). The splitting of each high-order vortex happens spontaneously because of the intrinsic random phase fluctuation.

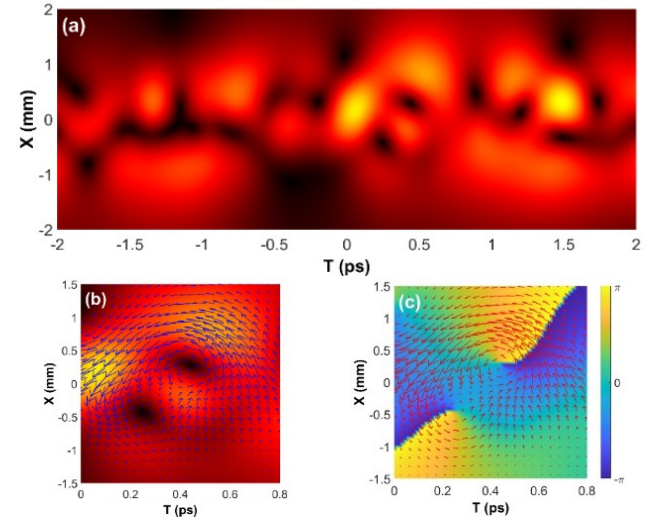

Fig.4. Partially coherent STOV with topological charge $l = 2$, $\sigma^2 = \pi^2$, (a) the train of multiple vortices; (b) amplitude and Poynting vectors and (c) phase and Poynting vectors show the vortex split.

The existence of the phase singularities of the partial temporal coherent STOV can be experimentally verified by the interference fringes which is the same method used to detect the coherent STOV [12]. Figure 5 shows the principle of the measurement. The reference and object beam are branched from the same partially coherent source, where the object is converted to a STOV. We can clearly see that the reference beam has multiple random peaks. In contrast, the corresponding peaks in the object beam tends to have amplitude dips due to the phase singularities. When the reference scans through the object, each peak of the reference will scan through the corresponding phase singularity in the object beam. Since the object and the reference have a high degree of spatial coherence within the coherent time, the time average of object-reference interference is to be captured by a CCD camera at a certain temporal delay. This detection method can not only prove the existence of the phase singularities, but also reveal the order of the topological charge. Mathematically, such a process can be expressed as:

$$\iint |E_o(x,y,t) + E_r(x,y,t+\Delta t)e^{j2\pi sin(\theta)y/\lambda}|^2 \, dt \quad (3)$$

where object $E_o$ and the reference $E_r$ have a time delay of $\Delta t$, merged with a small angle $\theta$, and $\lambda$ represents the central wavelength.

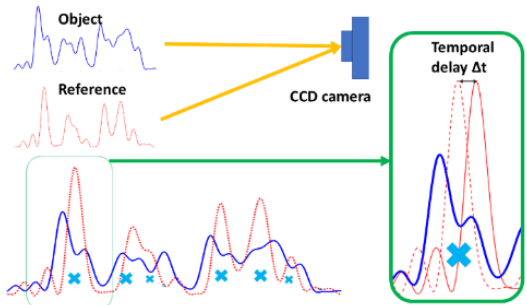

Fig.5. Principle of the measurement: fringe patterns are detected by the CCD camera. X-represent phase singularities in the object beam.

Figure 6 shows the numerically simulated fringe patterns for the partial temporal coherent STOV with topological charge $l = 1$. As the reference scans the object, we can observe the phase transition of the up and down fringes. At the center position, where the time delay is zero, we can see the π phase difference clearly in Figure 6(b). Since the spectral bandwidth is around 8 nm, the coherent time is around 450 fs. From Figure 6(a) to 6(c), the total time delay is around 85 fs, which is in the range of the coherent time.

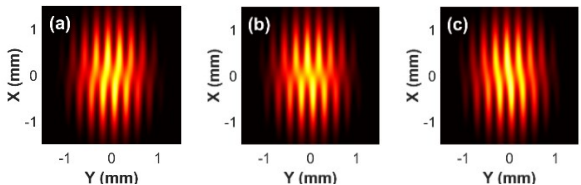

Fig.6. The theoretical interference fringe pattern of the partial temporal coherent STOV with topological charge $l$=1, (a) time delay=-43fs; (b) time delay=0fs; (c) time delay=43fs.

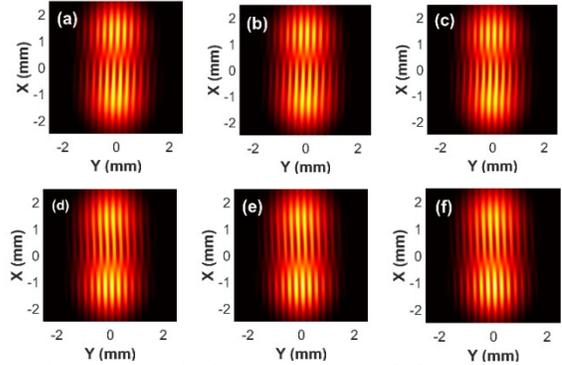

Fig.7. The theoretical phase transition of the partial temporal coherent STOV with topological charge $l$=2, (a) time delay=-119fs; (b) time delay=-93fs; (c) time delay=-68fs; (d) time delay=59fs; (e) time delay=85fs; (f) time delay=110fs.

The phase transition patterns also reveal the topological charge of the partial temporal coherent STOV, where we can see the π phase transition twice for topological charge l=2. From Fig.7 (a) to 7(c), we can see one transition occurs in the upper position, and from Fig.7 (d) to 7(f) is another transition that appears in the lower position.

The initial experiment was performed using amplified spontaneous emission (ASE) of a fiber laser. Figure 8 (a) shows the experimental setup which includes the pulse shaper with a 2D spatial light modulator for generating the partial temporal coherent STOV. An ytterbium-doped single-mode fiber laser was used for the experiment with a center wavelength of 1030 nm. The laser was driven below the lasing threshold to achieve ASE. It is noteworthy that the ASE from the single-mode fiber has a high degree of spatial coherence since only the fundamental transverse mode is available. Hence, it is reasonable to assume that the partial coherence exists only in the temporal domain. The spiral phase pattern projected onto the SLM will produce partial temporal coherent STOVs. Figure 8(b) shows the output spectrum of ASE with a FWHM of 7.8 nm bandwidth.

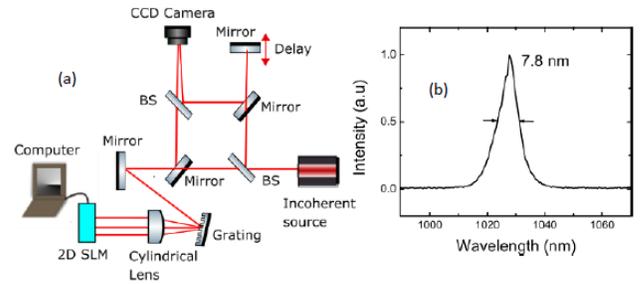

Fig. 8. (a) Schematic of the setup for partial temporal coherent STOV generation. BS-beam splitter (b) Spectrum of the ASE source.

Figure 9 shows the phase transition patterns observed by scanning the reference beam, for topological charges $l = +1, -1,$ and $+ 2$. Two phase transitions were observed for topological charge $l$=+2, which is in agreement with the theoretical prediction.

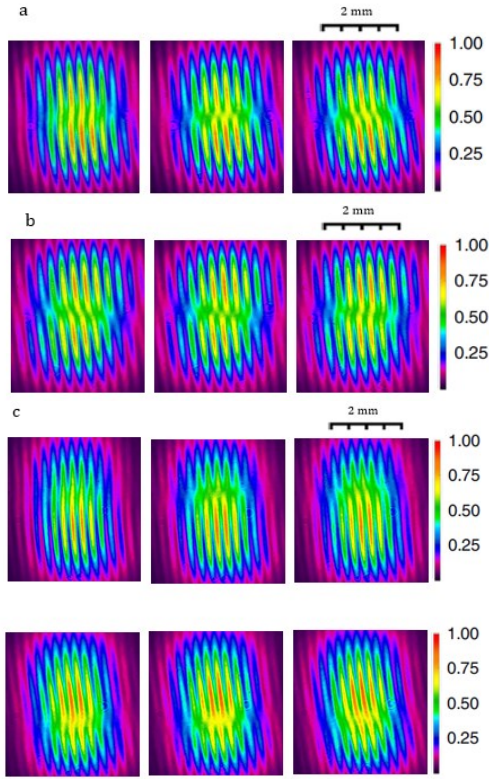

Fig. 9. Phase transitions for topological charge (a) $l$=+1, (b) $l$=-1, and $l$=+2. Time delays of the phase patterns from left to right are(a) -68 fs,0 fs and 32 fs (b) -113 fs, 0 fs and 50 fs (c) -216 fs,-131 fs,-103 fs,78 fs,131 fs and 180 fs respectively. In (a)-(c), color bars represent the intensities in arbitrary units.

Besides the ASE, a noise-like pulse (NLP) state from the fiber laser was also investigated since the NLP state provides a broad spectrum partial temporal coherent pulses where the phase relation between longitudinal modes is not strictly fixed [16].NLP states have higher peak power than ASE, hence, the STOV from the NLP state can be useful for high-power applications. Similar to the previous experiment, clear phase transitions were observed for topological charge $l$=+1 in the NLP state and they are shown in Figure 10.

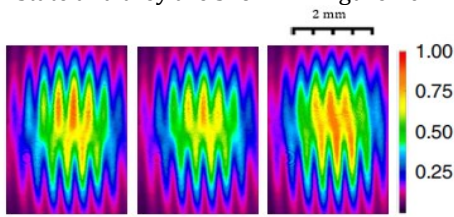

Fig. 10. Phase transition for topological charge l= 1 of the NLP state. Temporal delays to the left and right of the vortex center are 46 fs and 13 fs respectively.

In conclusion, we have successfully generated a partial temporal coherent STOV. The method implemented in our experiment will motivate the use of cheap broadband partially coherent sources for generating STOV. We believe that this partially coherent STOV will find new applications soon because it can be generated conveniently at a low cost.